\documentclass[reprint, superscriptaddress, amsmath, amssymb, aps, prb]{revtex4-2}
\usepackage{graphicx}
\usepackage{dcolumn}
\usepackage{bm}
\usepackage{placeins}
\usepackage[mathlines]{lineno}

\usepackage[usenames,dvipsnames]{xcolor}
 \usepackage{amsmath}
 \usepackage{amsfonts}
 \usepackage{amssymb}
 \usepackage[colorlinks=true,citecolor=blue,linkcolor=red]{hyperref}

 \usepackage{bbold}     
 \usepackage[makeroom]{cancel}  
 \usepackage{multirow}    
 \usepackage[normalem]{ulem}        
 \usepackage{array}
 \usepackage{pdfpages}
\makeatletter
 \AtBeginDocument{\let\LS@rot\@undefined}
 \makeatother

\begin{document}

\title{Suppressing Andreev bound state zero bias peaks using a strongly dissipative lead}

\author{Shan Zhang}
 \email{equal contribution}
\affiliation{State Key Laboratory of Low Dimensional Quantum Physics, Department of Physics, Tsinghua University, Beijing 100084, China}

\author{Zhichuan Wang}
 \email{equal contribution}
\affiliation{Beijing National Laboratory for Condensed Matter Physics, Institute of Physics, Chinese Academy of Sciences, Beijing 100190, China}

\author{Dong Pan}
 \email{equal contribution}
\affiliation{State Key Laboratory of Superlattices and Microstructures, Institute of Semiconductors, Chinese Academy of Sciences, P. O. Box 912, Beijing 100083, China}
\affiliation{Beijing Academy of Quantum Information Sciences, 100193 Beijing, China}

\author{Hangzhe Li}
\affiliation{State Key Laboratory of Low Dimensional Quantum Physics, Department of Physics, Tsinghua University, Beijing 100084, China}

\author{Shuai Lu}
\affiliation{State Key Laboratory of Low Dimensional Quantum Physics, Department of Physics, Tsinghua University, Beijing 100084, China}

\author{Zonglin Li}
\affiliation{State Key Laboratory of Low Dimensional Quantum Physics, Department of Physics, Tsinghua University, Beijing 100084, China}

\author{Gu Zhang}
\affiliation{Beijing Academy of Quantum Information Sciences, 100193 Beijing, China}

\author{Donghao Liu}
\affiliation{State Key Laboratory of Low Dimensional Quantum Physics, Department of Physics, Tsinghua University, Beijing 100084, China}

\author{Zhan Cao}
\affiliation{Beijing Academy of Quantum Information Sciences, 100193 Beijing, China}

\author{Lei Liu}
\affiliation{State Key Laboratory of Superlattices and Microstructures, Institute of Semiconductors, Chinese Academy of Sciences, P. O. Box 912, Beijing 100083, China}

\author{Lianjun Wen}
\affiliation{State Key Laboratory of Superlattices and Microstructures, Institute of Semiconductors, Chinese Academy of Sciences, P. O. Box 912, Beijing 100083, China}

\author{Dunyuan Liao}
\affiliation{State Key Laboratory of Superlattices and Microstructures, Institute of Semiconductors, Chinese Academy of Sciences, P. O. Box 912, Beijing 100083, China}

\author{Ran Zhuo}
\affiliation{State Key Laboratory of Superlattices and Microstructures, Institute of Semiconductors, Chinese Academy of Sciences, P. O. Box 912, Beijing 100083, China}

\author{Runan Shang}
\affiliation{Beijing Academy of Quantum Information Sciences, 100193 Beijing, China}

\author{Dong E Liu}
\email{dongeliu@mail.tsinghua.edu.cn}
\affiliation{State Key Laboratory of Low Dimensional Quantum Physics, Department of Physics, Tsinghua University, Beijing 100084, China}
\affiliation{Beijing Academy of Quantum Information Sciences, 100193 Beijing, China}
\affiliation{Frontier Science Center for Quantum Information, 100084 Beijing, China}

\author{Jianhua Zhao}
 \email{jhzhao@semi.ac.cn}
\affiliation{State Key Laboratory of Superlattices and Microstructures, Institute of Semiconductors, Chinese Academy of Sciences, P. O. Box 912, Beijing 100083, China}

\author{Hao Zhang}
\email{hzquantum@mail.tsinghua.edu.cn}
\affiliation{State Key Laboratory of Low Dimensional Quantum Physics, Department of Physics, Tsinghua University, Beijing 100084, China}
\affiliation{Beijing Academy of Quantum Information Sciences, 100193 Beijing, China}
\affiliation{Frontier Science Center for Quantum Information, 100084 Beijing, China}


\begin{abstract}
Hybrid semiconductor-superconductor nanowires are predicted to host Majorana zero modes, manifested as zero-bias peaks (ZBPs) in tunneling conductance. ZBPs alone, however, are not sufficient evidence due to the ubiquitous presence of Andreev bound states in the same system. Here, we implement a strongly resistive normal lead in our InAs-Al nanowire devices and show that most of the expected ZBPs, corresponding to zero-energy Andreev bound states, can be suppressed, a phenomenon known as environmental Coulomb blockade. Our result is the first experimental demonstration of this dissipative interaction effect on Andreev bound states and can serve as a possible filter to narrow down ZBP phase diagram in future Majorana searches. 
\end{abstract}

\maketitle

Electron tunneling in a dissipative environment has been widely studied before by probing the transport properties of a single nanoscale tunnel junction with highly resistive source/drain leads \cite{Delsing_1989, Devoret_1990, Cleland_1990, Ingold, Flensberg_1992, Joyez_1998, Zheng_1998}. The tunneling electrons interact with the electro-magentic plasmon modes of this ohmic environmental bath, resulting in a suppression of the tunneling current or conductance ($G\equiv dI/dV$) at low bias voltage ($V$) and temperature ($T$), a phenomenon known as `environmental Coulomb blockade' (ECB) or `dynamical Coulomb blockade'. This suppression exhibits a power-law as $G\propto$ max$(k_BT, eV)^{2r}$, similar to the tunneling behavior in a Luttinger liquid \cite{Chang_RMP}, where $k_B$ is the Boltzmann constant, $r = R/(h/e^2)$ is the ratio between the lead resistance (R) and the quantum resistance ($h/e^2$). $R$ or $r$ defines the dissipation strength. Though no real Luttinger liquid is present in dissipative tunneling, theory \cite{Safi_2004} has demonstrated a mapping connecting this two physically distinct systems where in both cases the tunneling electron's charge interacts with a continuum (linear) spectrum of bosonic modes. Later on, replacing the metallic tunnel junction with a semiconductor nanostructure, e.g. a quantum point contact \cite{Jezouin_2013,Pierre_PRX} or a quantum dot \cite{Gleb_PRB, Gleb_Nature, Gleb_NaturePhysics, Dong_PRB2014}, significantly increases the system's tunability and enables various quantum phase transitions. For example, in a quantum dot system with a single dot level resonantly coupled to the dissipative source and drain leads with the coupling strength being $\Gamma_{S/D}$, both experiment \cite{Gleb_Nature} and theory \cite{Dong_PRB2014} have demonstrated that 1) in the asymmetric coupling regime ($\Gamma_S \neq \Gamma_D$), the Coulomb conductance peaks are significantly suppressed as $T$ decreases, same to the typical ECB suppression for the single barrier case; 2) in the symmetric coupling regime ($\Gamma_S = \Gamma_D$), the Coulomb peak's height will, on the contrary, increase as $T$ decreases and finally saturates to the quantized conductance of $e^2/h$.      

Motivated by this striking contrast between symmetric regime versus asymmetric, theory \cite{Dong_PRL2013} has proposed that a strongly dissipative lead can be implemented in hybrid semiconductor-superconductor nanowire devices to distinguish signatures of Majorana zero mode (MZMs) \cite{Lutchyn2010, Oreg2010, Lutchyn2018Review, Prada2020}. Probing MZMs with a regular normal lead reveals zero-bias peaks (ZBPs) in tunneling conductance. The quantitative behavior of the ZBPs observed in experiments \cite{Mourik, Deng2016, Gul2018, Nichele2017, Zhang2021, Song2021}, however, does not fully follow the simplest MZM theory's prediction. For example, the perfect Majorana quantization \cite{DasSarma2001, Law2009} has not been observed yet. This quantization is enabled by Andreev reflection where the injected electrons and Andreev reflected holes `see' the same barrier twice, similar to the quantum dot case with symmetric resonant tunneling ($\Gamma_S = \Gamma_D$) which leads to perfect transmission and ZBP quantization. On the other hand, the biggest challenge of Majorana hunting comes from the co-existence of Andreev bound states (ABS) in the same device system \cite{Silvano2014}. These ABS-induced ZBPs, superficially similar to MZM signatures, can easily emerge due to potential inhomogeneity \cite{Prada2012,BrouwerSmooth, Liu2017, CaoZhanPRL} or disorder \cite{Patrick_Lee_disorder_2012, DEL-Disorder2018, GoodBadUgly, DasSarma2021above} which is currently unavoidable \cite{DasSarma_estimate}. The tunneling conductance of these ABS is also mediated by Andreev reflection where the tunnel couplings of the electrons and holes are usually different, similar to the quantum dot case with asymmetric coupling ($\Gamma_S \neq \Gamma_D$). As a result, the heights of these ZBPs are generally not quantized. Therefore, a dissipative lead with large enough $r$ could suppress the ubiquitous ABS-induced ZBPs (asymmetric coupling) at low $T$, while the quantized ZBPs due to symmetric coupling should still survive as long as the dissipation strength $r$ is less than 1/2 \cite{Dong_PRL2013}. 

In this paper, we engineer such a dissipative lead in our hybrid InAs-Al devices and demonstrate that strong dissipation can indeed suppress most of ZBPs formed from zero-energy ABSs. The survival of quantized ZBPs, possibly due to MZMs or quasi-MZMs, will be studied in the future. We note that the key idea of this proposal \cite{Dong_PRL2013} is based on the local features of Majorana resonance. Therefore, it can not distinguish topological MZMs from quasi-MZMs due to fine-tuned smooth potential \cite{BrouwerSmooth, WimmerQuasi, TudorQuasi} or disorder \cite{DasSarma2021Disorder}, which gives two spatially separated MZMs in the topological trivial regime.

Fig. 1a shows a scanning electron microscope (SEM) image of the device (Device A). The Cr/Au film (red) was made thin and resistive, serving as an ohmic dissipative environment (referred as `dissipative resistor' for the rest). This dissipative resistor is connected to the InAs-Al hybrid nanowire through the Ti/Au contacting lead (yellow). The resistance of the contacts is negligible due to its large thickness ($\sim$ 75 nm). The resistance of the dissipative resistor is estimated to be $\sim$ 5.46 k$\Omega$ based on an independent calibration of a four-terminal Cr/Au film fabricated together with this device (see Sfig. 1 for details). A total bias voltage $V_{bias}$ is applied to the left most yellow lead. The current $I$, after flowing through the dissipative resistor and the InAs-Al device, is drained and measured at the right most contact. Differential conductance $G \equiv dI/dV$ is then calculated in this two-terminal set up by subtracting the series resistance, $R_{series}$, contributed by the dissipative resistor (5.46 k$\Omega$) and fridge filters (see Sfig. 1). The bias $V$ across the InAs-Al device is calibrated by subtracting the bias drop shared by $R_{series}$: $V=V_{bias}-I\times R_{series}$. Therefore, $G=\frac{1}{dV/dI}=\frac{1}{\frac{dV_{bias}}{dI}-R_{series}}$. The side tunnel gate voltage ($V_{TG}$) tunes the tunnel barrier of the InAs, while the global back gate voltage ($V_{BG}$) tunes both the barrier and the superconducting part of the nanowire. The dissipative resistor should be lithographically close to the InAs-Al device to guarantee significant ECB effect: if their physical distance was too far away, ECB would fade out and the dissipative resistor would obey Ohm's law like the fridge filters.

\begin{figure}[htb]
\includegraphics[width=\columnwidth]{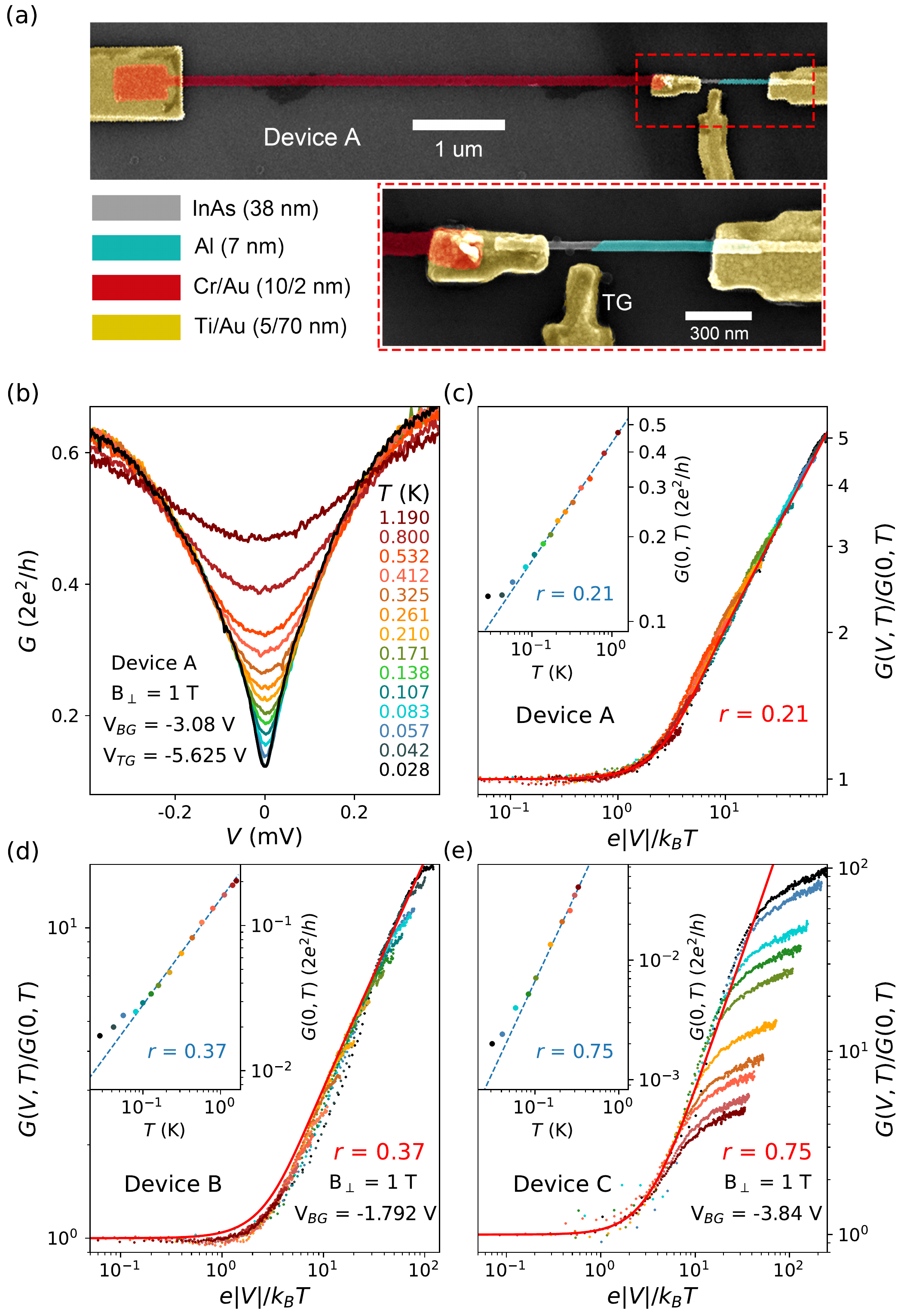}
\centering
\caption{(a) False-color SEM image of Device A. Thickness of the dissipative resistor (red), contact/gate (yellow) and InAs (diameter) are labeled in lower left. The substrate is p+ Si covered by 300 nm thick SiO$_2$, acting as a global back gate. (b) Differential conductance $G(V, T)$ of device A at different temperatures $T$. $B_{\perp}$ = 1 T. (c) All the curves in (b) (negative bias branch), re-plotted using dimensionless units, collapse onto a universal curve (red) with minor deviations. Inset, $T-$dependence of the zero-bias $G$. (d) and (e) for Device B and C with different dissipative resistors.}
\label{fig1}
\end{figure}

To demonstrate ECB in our device, we first apply an out-of-plane magnetic field ($B$) of 1 T (perpendicular to the Al film) to suppress its superconductivity. The device is then equivalent to a normal tunnel junction connected to a dissipative lead. Fig. 1b shows the significant suppression of $G$ at low bias $V$ and temperature $T$, consistent with the hallmark of ECB (`dissipative tunneling'). In Fig. 1c, we re-scale all the curves in Fig. 1b (the negative bias branch) using dimensionless units, where they all collapse onto a single universal curve (the red line in Fig. 1c) with minor deviations. The red line is obtained by performing numerical derivative of the expression for the dissipative tunneling of a single tunnel barrier \cite{Gleb_Nature}: $I(V,T) \propto VT^{2r}|\Gamma (r+1+ieV/2\pi k_BT)/\Gamma (1+ieV/2\pi k_BT)|^2$ where $k_B$ is the Boltzmann constant and $\Gamma$ the Gamma function. $r$ is the dissipation strength extracted from the power-law exponent, $G\propto T^{2r}$, of the $T$-dependence for the zero-bias $G$ (Fig. 1c inset).  For $T<$ 100 mK, $G$ deviates from the power-law (blue line), suggesting a gradual saturation of the electron $T$: fridge $T$ of $\sim$ 30 mK roughly corresponds to an electron $T\sim$ 50 mK. We used the electron $T$, estimated based on this `power-law' thermometer \cite{HenokThesis}, for the re-scaled x-axis $e|V|/k_BT$ in Fig. 1c for curves of $T<$ 100 mK.

\begin{figure*}[htb]
\includegraphics[width=0.75\textwidth]{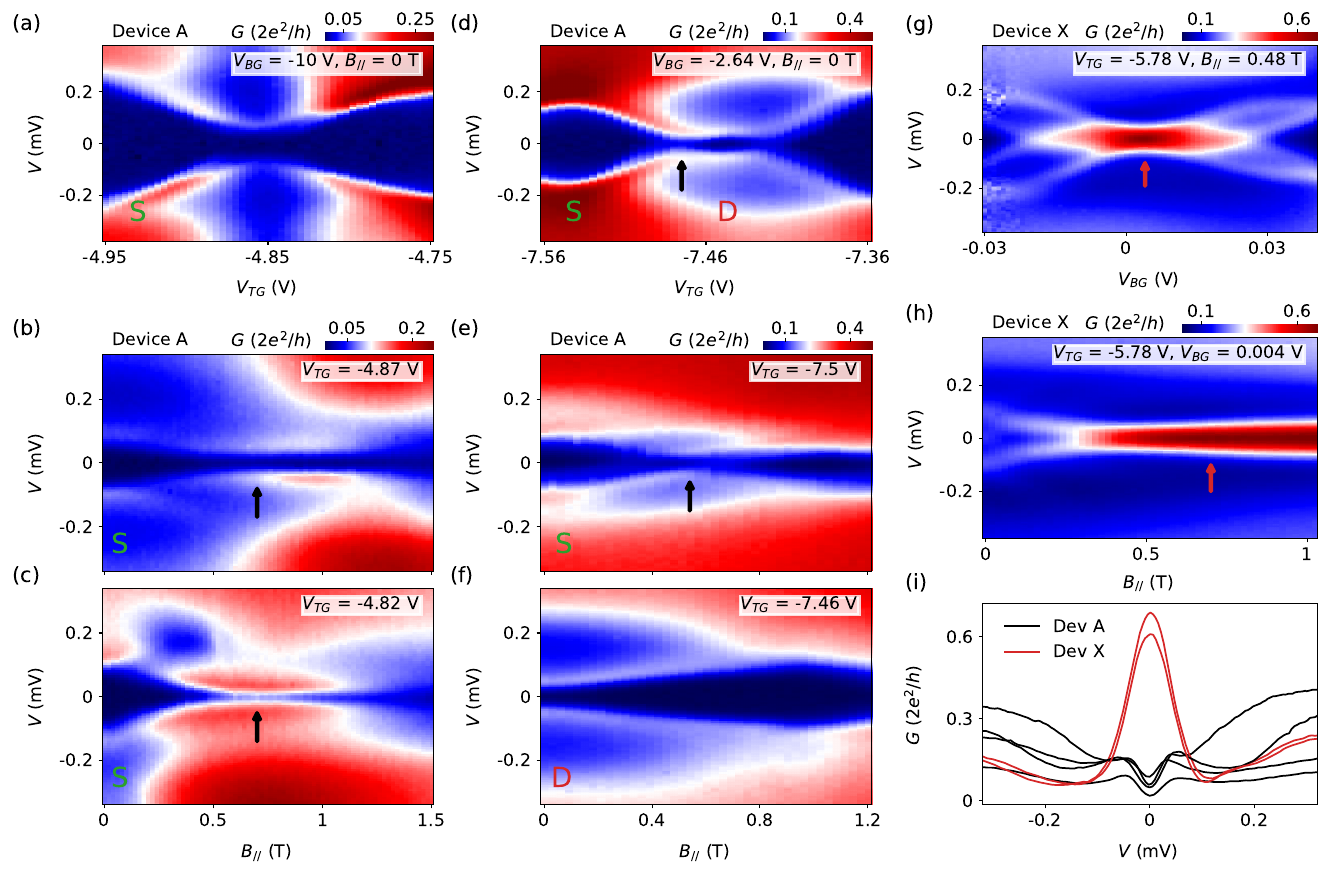}
\centering
\caption{(a)-(f) are from Device A. (a) $V_{TG}$ scan of a singlet-ABS at $B$ = 0 T. (b-c) $B$-scan of the ABS from (a) ($V_{TG}$ labeled). (d) $V_{TG}$ scan of another ABS showing singlet-doublet ground state switching at $B=0$ T. (e-f) $B$-scan of the singlet (S) and doublet (D) ABS. (g-h) Gate and $B$ scans of an ABS in Device X, a regular InAs-Al device without the dissipation resistor. (i) Black curves are line-cuts from (b-e) (black arrows) and red curves are line-cuts from (g-h), labeled with red arrows. $B$ aligned with the nanowire axis, with a maximum field value of 1.02 T due to its orientation. $T\sim$  20 mK.}
\label{fig2}
\end{figure*}

The dissipation strength, $r = 0.21$, translates to an effective dissipation resistance $r \times h/e^2=5.42$ k$\Omega$, roughly consistent with our independent estimation of the dissipative resistor $\sim$ 5.46 k$\Omega$. We have also checked similar power-law behaviors in Device B and C, designed with larger dissipative resistors as shown in Fig. 1d and Fig. 1e. The extracted exponent $r = 0.37$ (0.75) for Device B (C) corresponds to an effective dissipation resistance of 9.55 (19.36) k$\Omega$, also roughly consistent with our independent estimation of 7.5 (27.28) k$\Omega$. The deviations might be due to inaccurate estimations of the dissipative resistor (see Sfig. 1) or contributions from other dissipation sources. When the dissipative resistor's $R$ is comparable to the resistance of the InAs-Al tunneling junction ($R_T$), the effective dissipative resistance should be replaced by $1/(1/R+1/R_T)$ \cite{Flensberg_1991}, also causing sizable deviations. Power-law fits at the other gate voltages (see Sfig. 2) show fluctuations possibly due to reasons mentioned above. In Fig. 1e, we note that $G$ at larger bias deviates from the universal red line. This deviation, also expected for Device A$\&$B if the bias was large enough, is because that the power-law is only applicable within a finite energy bandwidth \cite{Zheng_1998}. This cut-off energy/frequency is set by $1/RC$ where $C$ is the junction capacitance which further determines the charging energy of the junction. 
 
We now set $B$ back to zero and study the superconducting behavior in a dissipative environment. In the literature of `Majorana nanowires', the term `dissipation' usually refers to disorder at the InAs-Al interface which causes soft induced superconducting gaps \cite{Takei2013,LiuDissipation}. To avoid confusion, in this paper we refer `dissipation' only to effects caused by the dissipative resistor. In fact for our InAs-Al nanowires (without dissipation), we have shown the atomically abrupt InAs-Al interface \cite{Pan2020}, hard induced gaps and large ZBPs \cite{Song2021}. Now with the dissipative resistor (Device A), the gap, co-exists with Coulomb blockade, still remains hard (see Sfig. 3). But the coherence peaks are smeared, possibly due to non-equilibrium dissipation. Nevertheless, we can still resolve clean Andreev bound states (ABS) as shown in Fig. 2.

Fig. 2a shows the gate dependence of an ABS in Device A. The two sub-gap levels do not cross, suggesting a singlet ground state \cite{Silvano2014}. As a result, a magnetic field ($B$, aligned with the nanowire) Zeeman splits the two sub-gap peaks and drives the inner peaks towards zero as shown in Fig. 2bc (the splitting and the outer peaks are sometimes barely visible). In regular devices without a dissipative lead, further increasing $B$ can lead to level crossing and ZBP formation as shown in Fig. 2h for Device X. This ZBP due to ABS level crossing is a ubiquitous observation and well established phenomenon \cite{Silvano2014, Pan2020}. Now back to our Device A with dissipation, Fig. 2bc show that the `expected' ZBPs at the expected crossing points (arrows) are suppressed to split-peaks (see Sfig. 4 `waterfall' plots of line-cuts). This suppression of ZBPs in devices with a strongly dissipative lead is the main observation of the paper. We attribute this observation to the interaction effect (ECB) caused by the dissipative environment. ECB suppresses $G$ in a non-uniform way: the suppression is stronger for lower $T$ and energy (bias). Therefore, the zero-bias $G$ at base $T$ is suppressed the most, while the high-bias $G$ is less affected, making the expected ZBP to split-peaks \cite{Dong_PRL2013}. 

Fig. 2d shows another ABS case where the ground state can be continuously gate tuned from singlet (S) to doublet (D), see labeling. At the singlet-doublet switching point (black arrow), the sub-gap states are expected to cross and form a ZBP in regular devices without dissipation. Here with dissipation, the `expected' ZBP is again suppressed. Fig. 2ef show the $B$-dependence of the ABS. For the singlet case (Fig. 2e), the two peaks merge towards zero at the field (black arrow) where the `expected' ZBP is suppressed. In Fig. 2f, the two peaks move away from zero energy when increasing $B$, consistent with the doublet behavior \cite{Silvano2014}. For additional $B$- and gate-scans of the singlet and doublet ABSs in Fig. 2a-f, see Sfig. 5 and Sfig. 6. 

For comparison, we show a ZBP data set in control Device X (without the dissipative resistor) in Fig. 2gh. We have also applied a perpendicular $B$ of 1 T for Device X to suppress its superconductivity and test/confirm no-power law suppression as Fig. 1 (see Sfig. 7). The ZBP, formed from level merging, shows some robustness (non-split) in gate (Fig. 2g) and $B$ (Fig. 2h) scans. The ZBP is not quantized, ruling out the topological origin, see Sfig. 8 additional scans. In Fig. 2i we show line-cuts to contrast the suppressed ZBPs from Device A (black curves) and ZBPs from Device X (red curves).

We now study the $T$-dependence of the ABS with dissipation. Fig. 3a shows the ABS from Fig. 2d with $B=0.3$ T, measured at base $T$. The measurement was then repeated at different $T$s (see Sfig. 9). Fig. 3b shows a near-zero energy ABS line-cut from Fig. 3a for different $T$s (only four curves are plotted for clarity). The solid lines are theory simulations using the formula $G(V,T)=\int_{-\infty}^{+\infty}G(\epsilon,0) \frac{\partial f(eV-\epsilon, T)}{\partial \epsilon} d\epsilon$, where $f(E,T)=\frac{1}{e^{E/kT}+1}$ is the Fermi distribution function. For convenience, note the unit conversion between $G(V, T)$ and $G(\epsilon, 0)$ for $eV$ and $\epsilon$. We assume the $G(V, T=20$ mK$)$ curve as $G(V, 0)$ to calculate $G(V,T)$ for different $T$s which should be valid for $T$ much larger than 20 mK. The simulation, only taking thermal averaging effect into account, shows significant deviations from data, suggesting that thermal averaging alone can not explain the $T$-dependence of this near zero-energy ABS. The measured $G$ larger than the thermal simulation suggests the lifting of ECB suppression at higher $T$s.

We further plot the zero-bias $G$ of this line-cut for all $T$s in Fig. 3c (black dots), and find a power-law behavior with an exponent of $1.8r$, close to $2r$ assuming $r$ = 0.21. At other gate voltages where the ABS has finite energies, we also find similar power-law-like behavior within an intermediate $T$-range (from $\sim$ 60 mK to $\sim$ 300 mK), see the brown and orange dots as two examples with exponents of $4.4r$ and $8r$. Fig. 3a (lower panel) shows the exponent at different gate voltages (in the left- and right-most regions without data, we can not find a reasonable power-law-like fit). For $T$-dependence of the singlet ABS (Fig. 2a-c) and more power-law-like fittings, see Sfig. 9. 

In a recent theory work \cite{Dong_2021}, we have studied dissipative tunneling of ABS mediated by Andreev reflection and showed that the power-law exponent can be $8r$ or $4r$, corresponding to coherent or incoherent Andreev reflection processes. Other intermediate exponent values between $8r$, $4r$ and $2r$ (corresponding to normal tunneling due to e.g. soft gap or gap-closing) could also be achieved depending on the mixture between different processes. Whether our observation here corresponds to the theory situations discussed above remains as an open question and requires further systematic study. For example, we note the $T$-range of the power-law in Fig. 3c for $8r$ is not large (only half a decade), therefore hard to be used as an exclusive tool for theory understanding. For $T <60$ mK, the data deviation is possibly due to the saturation of electron $T$. For $T>$ 300 mK, the deviation is probably due to the softening of the superconducting gap where incoherence Andreev reflection or normal tunneling due to quasi-particle poisoning starts to contribute. Nevertheless, optimizations are needed for better power-laws to fully understand `dissipative Andreev tunneling'.

\begin{figure}[tb]
\includegraphics[width=\columnwidth]{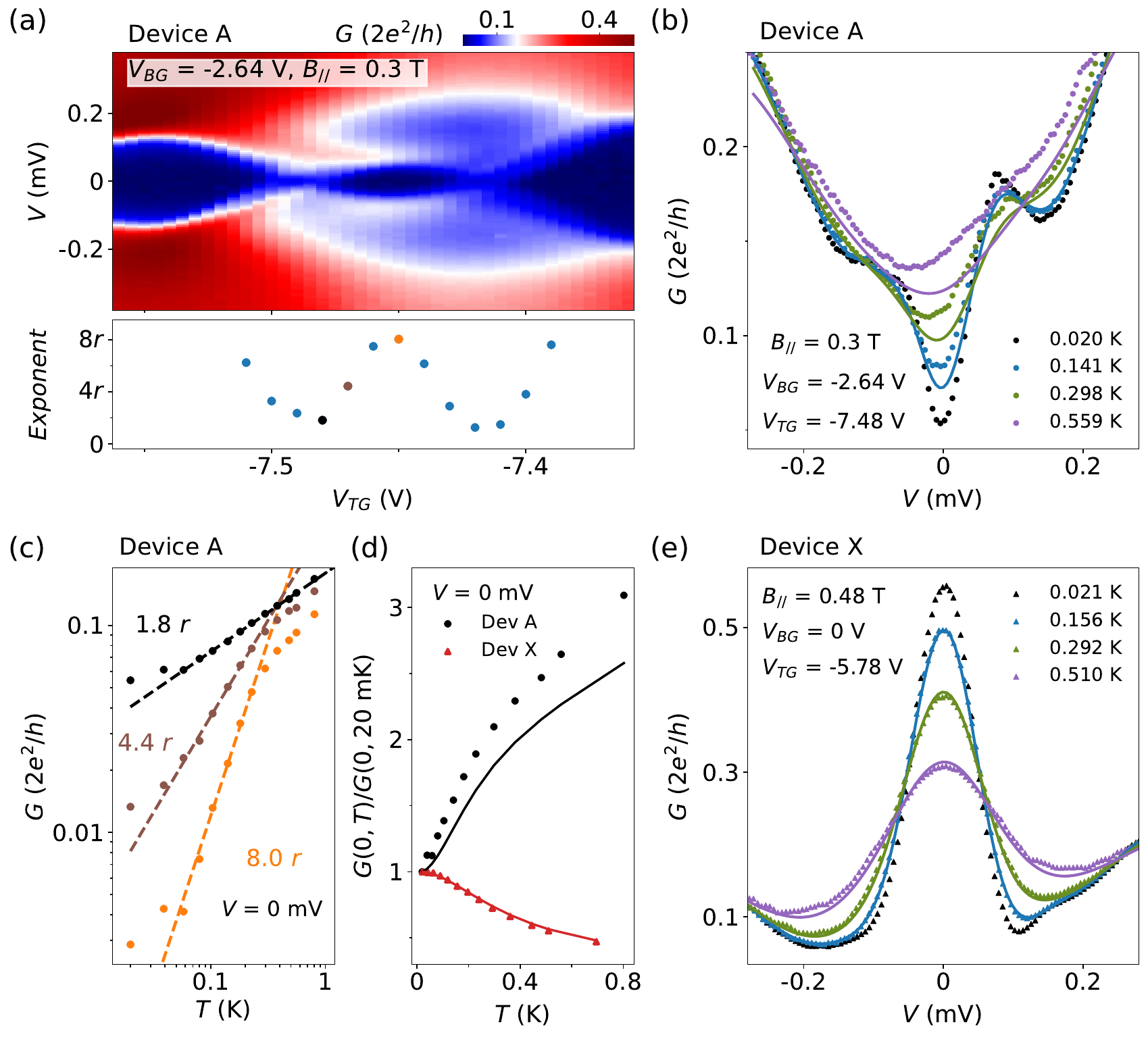}
\centering
\caption{(a) Upper, same with Fig. 2d but at $B=$0.3 T. $T\sim$ 20 mK. Lower, extracted power-law exponents from $T$-dependence in unit of $r$, assuming $r=0.21$. (b) $T$-dependence of a line-cut in (a), different $T$s shown as different colored dots. The corresponding colored lines are theory simulations assuming only thermal averaging effect. (c) $T$-dependence of the zero-bias $G$ from (a) at three $V_{TG}$s, see dots with corresponding color in lower panel (a). Dashed lines are the power-law fits. (d) $T$-dependence of the normalized zero-bias $G$ from (b) for Device A (black dots) and Device X (red dots), together with thermal averaging simulations (solid lines). (e) $T$-dependence of a ZBP (a line-cut from Fig. 2g) in control Device X (dots) and thermal averaging simulations (lines). }
\label{fig3}
\end{figure}

In Fig. 3d we re-plot the black dots from Fig. 3c and show the significant deviation from the thermal averaging simulation (black line). This deviation is also observable in Fig. 3b. By contrast, in our control Device X (without dissipation), the $T$-dependence of ZBP shows reasonable agreement with the simulation (without any fitting parameter) for both zero bias (Fig. 3d red) and finite bias (Fig. 3e), suggesting that thermal averaging is indeed the dominating effect in the $T$-dependence of zero-energy ABS in Device X. This sharp contrast between Fig. 3b and Fig. 3e confirms the Fermi liquid $T$-dependence for Device X (without dissipation) and suggests the non-Fermi liquid $T$-dependence in Device A (with dissipation). We note the softening of the superconducting gap at higher $T$, not taken into account in the simulation, could also cause small deviations between data and simulation in Device X especially for non-zero energy ABS (see SFig. 8). Therefore, we restrict our thermal averaging simulation to only zero or near-zero energy ABS.

\begin{figure}[tb]
\includegraphics[width=\columnwidth]{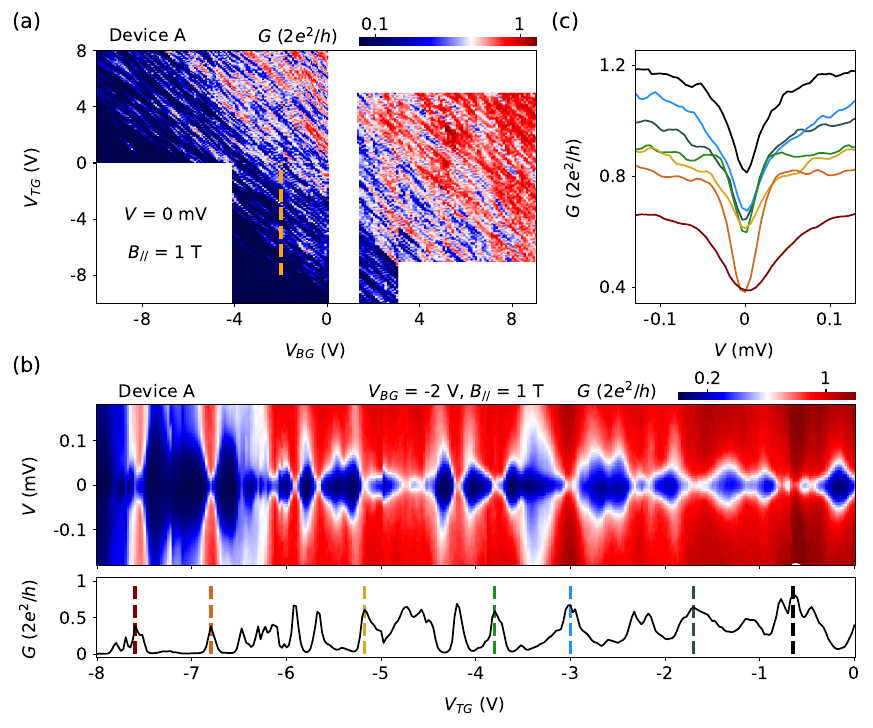}
\centering
\caption{(a) Zero-bias conductance map at $B=1$ T over a large $V_{BG}$ and $V_{TG}$ range. (b) $G$ versus $V_{TG}$ and $V$ with the zero-bias line-cut shown in the lower panel, corresponding to the orange dashed line in (a). (c) Vertical line-cuts from (b) labeled by the corresponding colored dashed lines, resolving no ZBPs. $T\sim$ 20 mK.}
\label{fig4}
\end{figure}

In Fig. 2 we have presented two examples of ABS to show that the expected ZBPs can be suppressed due to dissipation. In fact, similar behavior is observed in most of the parameter ranges explored in Device A. Fig. 4a shows the zero-bias $G$ of Device A over a large gate voltage range at a parallel field of 1 T where ZBPs are likely to occur \cite{Song2021}. The dense diagonal lines suggest formation of unintentional quantum dots and possibly ABSs. However, further bias scan (Fig. 4b) over a typical line-cut in Fig. 4a resolves no ZBP at all (see Fig. 4c line-cuts). We have also checked other regions of the parameter space in Device A, and did not find any clear ZBPs (see Sfig. 10). This absence of ZBP in almost all the explored parameter regions is dramatically different from our previous experience where without dissipation, ZBPs can be easily and routinely found \cite{Pan2020, Song2021}. Very rarely, very faint and small ZBP-like feature can be barely visible, see Sfig. 10 the only case we found in Device A. For Device B and C, we also did not find any ZBP (see Sfig. 3). We note this observation is not inconsistent with our expectation based on previous experience in regular devices where 1) non-quantized ZBPs are ubiquitous and can be easily found in every device \cite{Michiel2018}, therefore suppressed by dissipation; 2) large ZBPs near the quantized value (resolving plateau-like feature) are rare and not in every device \cite{Zhang2021, Song2021}. 

The dissipation resistance or strength $r$ is a crucial parameter. If $r$ is too strong (larger than 1/2), theory \cite{Dong_PRL2013} predicts that all ZBPs (including the ones due to Majoranas) will be suppressed. If $r$ is less than 1/2, Majorana or quasi-Majorana ZBPs could survive while trivial ZBPs will be suppressed (the latter is demonstrated in this paper). But this does not mean that $r$ can be arbitrarily small. For example, if $r$ is approaching zero, corresponding to the case without dissipation (Device X), apparently trivial ZBPs will revive. This weak or intermediate dissipative regime will be explored in a future separate work. We identify the $r$-value of 0.21 or larger in this paper as the strong dissipation regime based on the observation that most of ABS ZBPs can be suppressed. 

To summarize, we have implemented a strongly dissipative lead in our InAs-Al hybrid nanowire devices. The dissipative environment, confirmed by the observation of environmental Coulomb blockade, power-laws and non-Fermi liquid temperature dependence, can significantly suppress the Andreev bound state zero bias peaks, which ubiquitously exist in hybrid nanowires and disturb Majorana detections, see also \cite{Dong_2021} our recent related theory work. Our device set-up could serve as a possible filter to narrow down the ZBP phase diagram in future Majorana searches \cite{Dong_PRL2013, Dong_PRB2020, Gu_2020, NextSteps}.

\textbf{Acknowledgment} We thank Gleb Finkelstein, Harold Baranger and Chung-Ting Ke for valuable discussions. Raw data and processing codes within this paper are available at https://doi.org/10.5281/zenodo.5630076. This work is supported by Tsinghua University Initiative Scientific Research Program, National Natural Science Foundation of China (Grant Nos. 11974198, 92065106 $\&$ 61974138), Beijing Natural Science Foundation (Grant No. 1192017). D. P. acknowledges the support from Youth Innovation Promotion Association, Chinese Academy of Sciences (No. 2017156).

\bibliography{mybibfile}


\onecolumngrid

\clearpage



\section*{Supplementary material}

# Supplement Information: Suppressing Andreev bound state zero bias peaks using a strongly dissipative lead

Shan Zhang,[1, *] Zhichuan Wang,[2, *] Dong Pan,[3, 4, *] Hangzhe Li,[1] Shuai Lu,[1] Zonglin
Li,[1] Gu Zhang,[4] Donghao Liu,[1] Zhan Cao,[4] Lei Liu,[3] Lianjun Wen,[3] Dunyuan Liao,[3]
Ran Zhuo,[3] Runan Shang,[4] Dong E Liu,[1, 4, 5, †] Jianhua Zhao,[3, ‡] and Hao Zhang[1, 4, 5, §]

[1]*State Key Laboratory of Low Dimensional Quantum Physics,
Department of Physics, Tsinghua University, Beijing 100084, China*
[2]*Beijing National Laboratory for Condensed Matter Physics,
Institute of Physics, Chinese Academy of Sciences, Beijing 100190, China*
[3]*State Key Laboratory of Superlattices and Microstructures, Institute of Semiconductors,
Chinese Academy of Sciences, P. O. Box 912, Beijing 100083, China*
[4]*Beijing Academy of Quantum Information Sciences, 100193 Beijing, China*
[5]*Frontier Science Center for Quantum Information, 100084 Beijing, China*

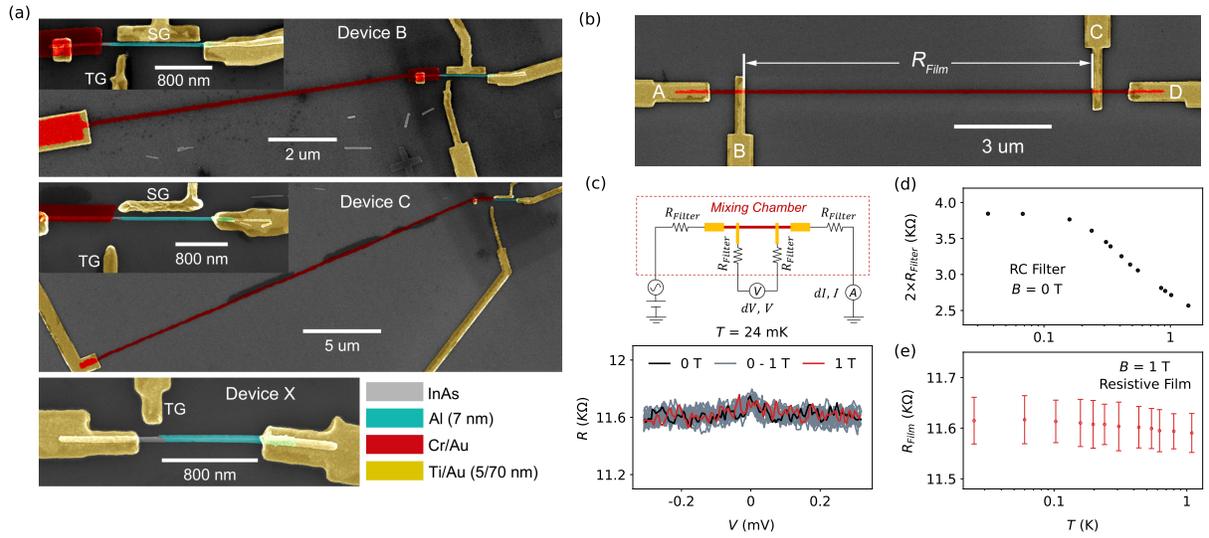

Supplement Fig. 1: Device SEM images, dissipative resistor and fridge filter calibration. (a) Upper, middle and lower panels are the SEM images for Device B, C and X. Insets of upper and middle panels show the zoom-in of the InAs-Al parts. Device X does not have the dissipative resistor, acting as a control device. The fabrication sequence for Device A-C is: 1) wet-etching of Al film, 2) Cr/Au dissipative resistor, 3) Ti/Au contacts and gates. The total thickness of the Cr/Au film for Device B (C) is 14.3 (12.6) nm. (b) False-color SEM of a Cr/Au dissipative film (red) fabricated together with the dissipative resistor of Device A on the same substrate chip. The four electrode contacts (yellow) are Ti/Au, fabricated together with the contacts of Device A. (c) Upper panel shows the four-terminal measurement set up: the current $I$ flows from A to D and the voltage $V$ is measured between B and C. Each fridge line has a RC-filter on the mixing chamber plate (see $R_{Filter}$ in the schematic, capacitors not shown). Differential resistance of the film $R_{Film} = dV/dI$ is shown in the lower panel without obvious dependency on bias $V$ and magnetic field ($B$), measured at base $T$. (d) Independence measurement of two fridge filters and its $T$-dependence, by shorting the two corresponding fridge lines at mixing chamber. (e) $T$-dependence of $R_{Film}$. Square resistance is extracted for this dissipative film which is further used to estimate the resistance of the dissipative resistor for Device A. For Device B and C, belonging to different fabrication rounds, we do not have this four-terminal measurement and therefore estimate the resistance of dissipative resistors based on its film thickness. The accuracy, in this way, is less and may cause deviations from the power-law exponents as shown in Fig. 1de. All the reported $G$ refers to only the InAs-Al part where the resistance of the dissipative resistor and filters is excluded. The bias voltage drop $V$ also only refers to the InAs-Al part, with the contribution from the dissipative resistor and $R_{Filter}$ subtracted.

--------
* equal contribution
† dongeliu@mail.tsinghua.edu.cn
‡ jhzhao@semi.ac.cn
§ hzquantum@mail.tsinghua.edu.cn

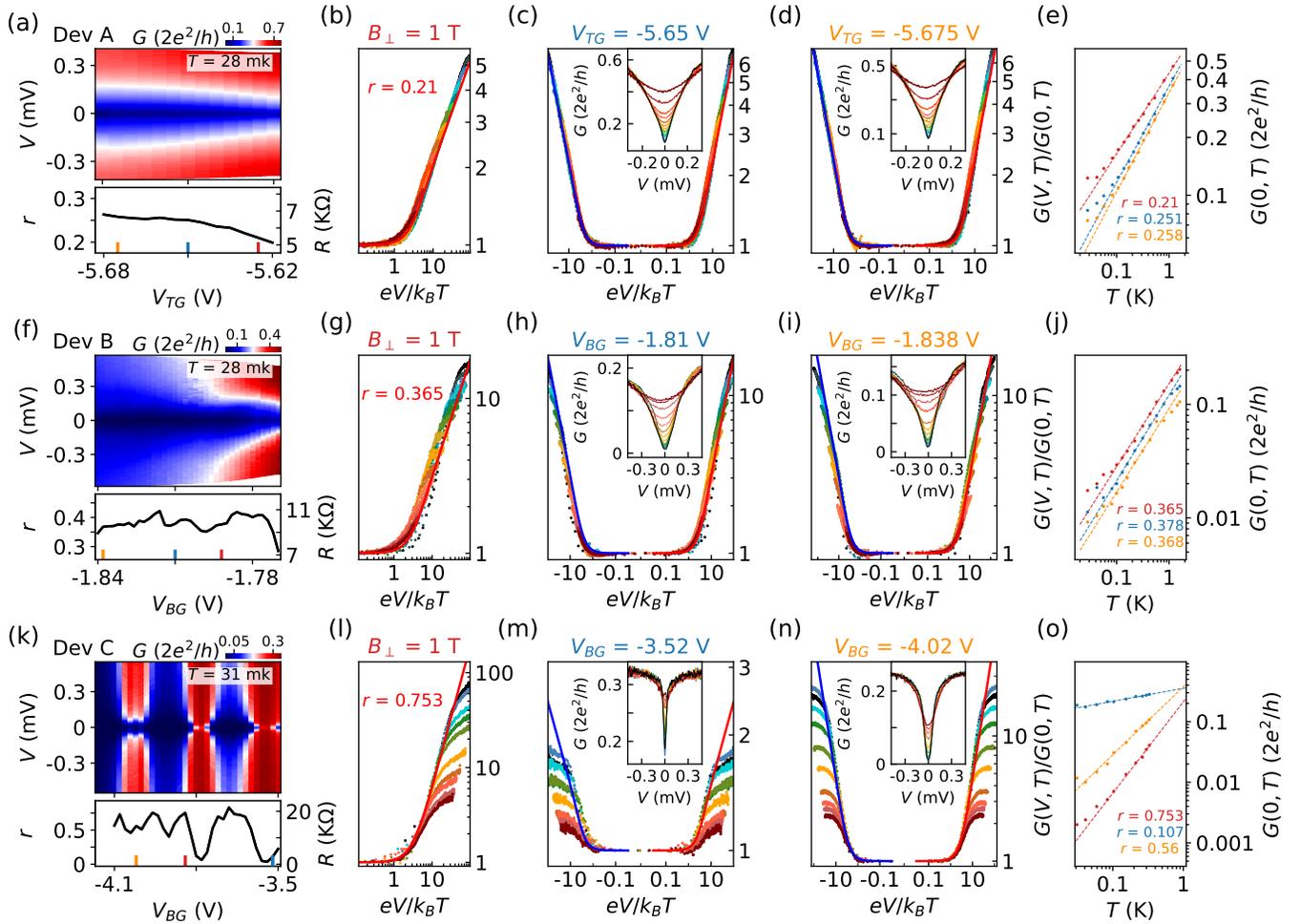

Supplement Fig. 2: More on power-law calibrations for Device A (a-e), B (f-j) and C (k-o). (a) $G$ vs $V$ and $V_{TG}$ at $B_\perp = 1$ T. $V_{BG} = -3.08$ V. $T = 28$ mK. Same measurement is repeated from 28 mK to 1.19 K (not shown). (b) The positive bias branch of Fig. 1bc, corresponding to a vertical line-cut in (a) at $V_{TG} = -5.625$ V. (c-d) $T$-dependence of two more line-cuts ($V_{TG}$ labeled), in linear (inset) and log scale (using dimensionless units). The x-axis is in linear-scale between -0.1 and 0.1 and log-scale outside this range. The colors correspond to different $T$s (see Fig. 1b). The blue and red lines are the universal curves with $r$ extracted from (e). (e) The $T$-dependence of the zero-bias $G$ for these three $V_{TG}$-values, resolving power-laws. The low-$T$ deviation is due to the saturation of the electron $T$. The extracted exponent fluctuates for different $V_{TG}$s with possible reasons discussed in the maintext. The lower-panel of (a) shows this power-law exponent for all the measured $V_{TG}$-range. (f-j) and (k-o) same with (a-e) but for Device B and C, respectively. The power-law fits in (m), corresponding to a point close to the Coulomb degeneracy, shows sizable deviations. This is possibly because that near the degeneracy point, the InAs-Al part could no longer be treated as a 'featureless' single barrier.

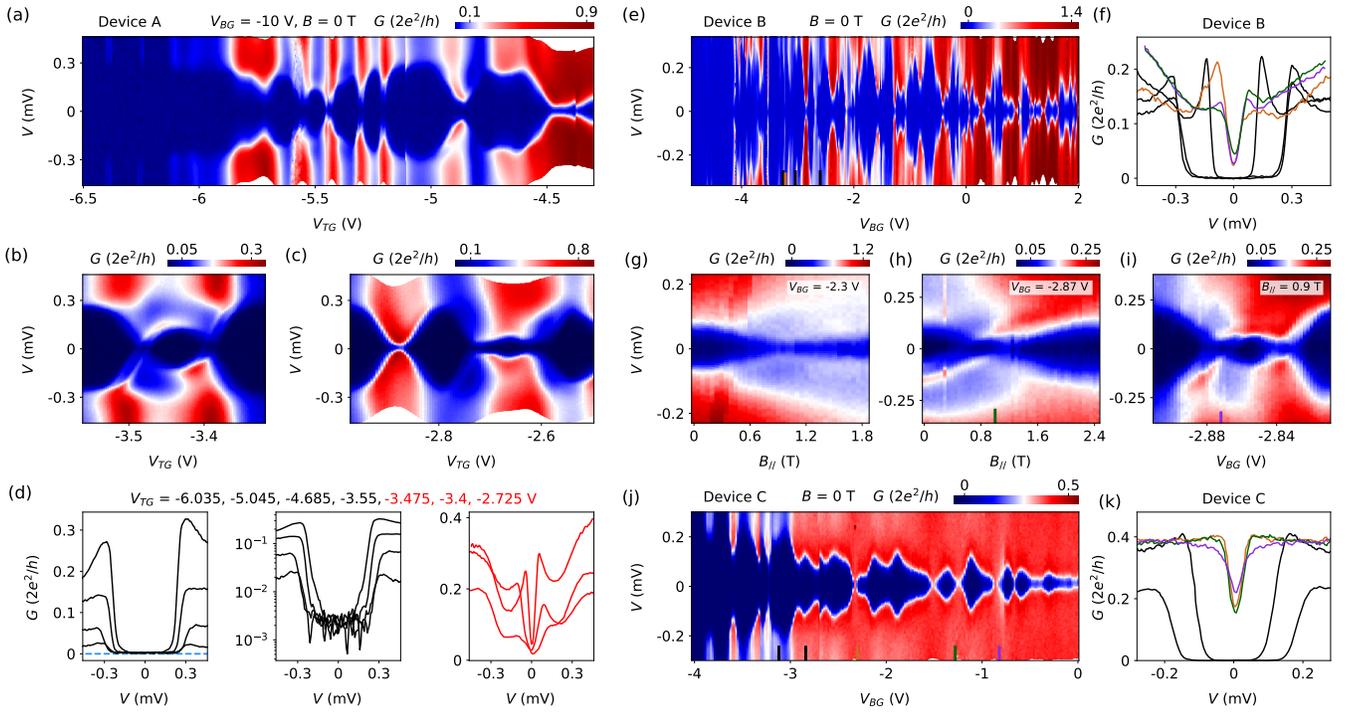

Supplement Fig. 3: (a-c) Basic transport characterization of Device A at $B = 0$ T at three different $V_{TG}$ scanning ranges. No ZBPs are observed. (d) Line-cuts from (a-c): left and middle panels show the superconducting gap (with possible Coulomb blockade) in linear and logarithmic scales while the right panel shows line-cuts where 'expected' ZBPs due to level crossing are suppressed. (e) Basic transport characterization of Device B at $B = 0$ T. No ZBPs are observed. (f) Line-cuts of the superconducting gap (black), possibly affected by Coulomb blockade, and sub-gap states (colored). (g-h) $B$-scans of Device B at two different gate settings, where $B$-induced level crossings are observed. The expected ZBPs are suppressed. (i) Gate scan of Device B where gate induced level crossings and suppressed ZBPs are observed. Line-cuts from (h-i) at the expected crossing points are shown in (f) (with corresponding colors). (j) Basic transport characterization of Device C at $B = 0$ T with line-cuts shown in (k). We note that the superconducting coherence peaks are smeared out as the dissipation strength gets stronger (from Device A to C). Moreover, the existence of unintentional quantum dots and Coulomb blockades also affects the gap measurements. $T \sim 20$ mK for all panels.

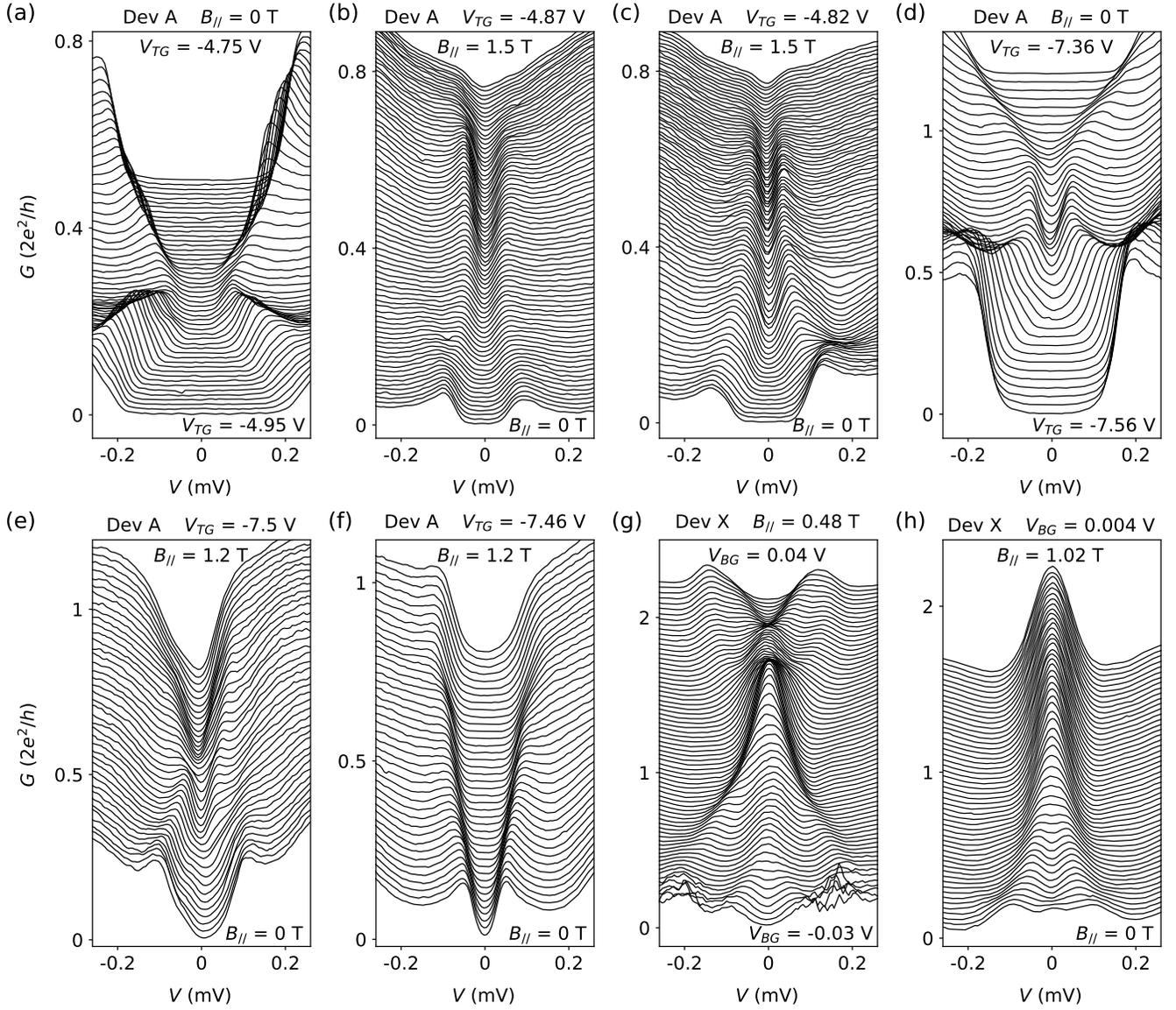

Supplement Fig. 4: Waterfall plots of Fig. 2a-h. Vertical offsets (for clarity) are 0.01 for panel a-c, 0.02 for panel e, f and 0.03 for panel d, g h, all in unit of $2e^2/h$. The suppressed ZBPs at the expected crossing points in panel b-e for Device A and ZBPs in panel g-h for Device X are clearly visible.

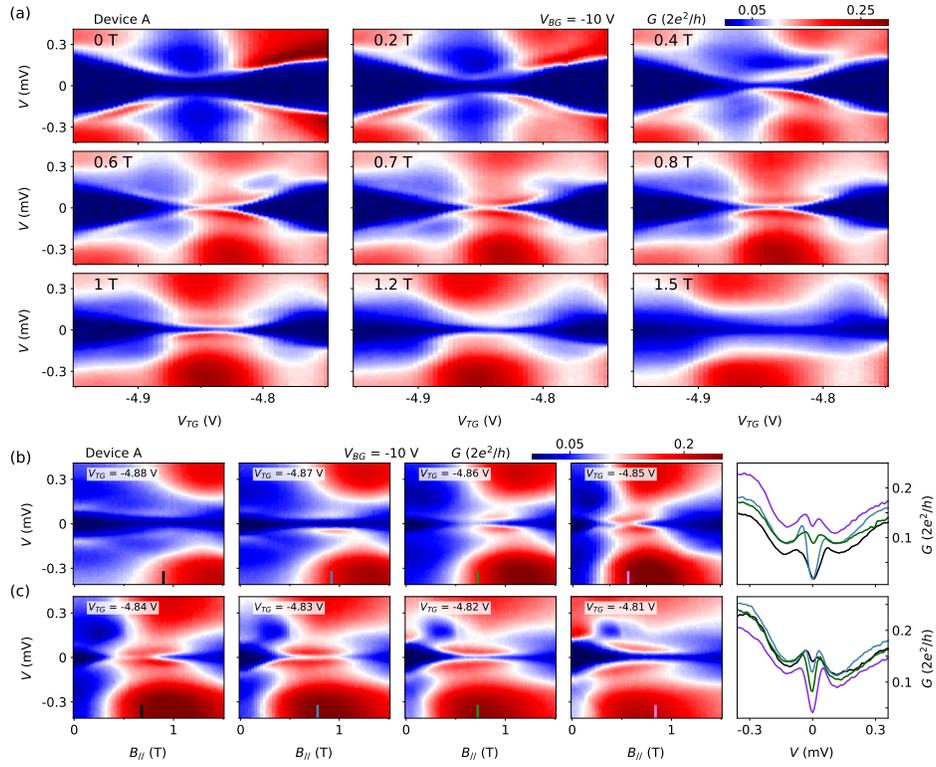

Supplement Fig. 5: Additional scans of the ABS in Fig. 2a-c. T ∼ 20 mK. (a) $V_{TG}$ scans ($B$ labeled). (b-c) $B$ scans ($V_{TG}$ labeled), and line-cuts of suppressed ZBPs (right) with corresponding color bars labeled in the left panels.

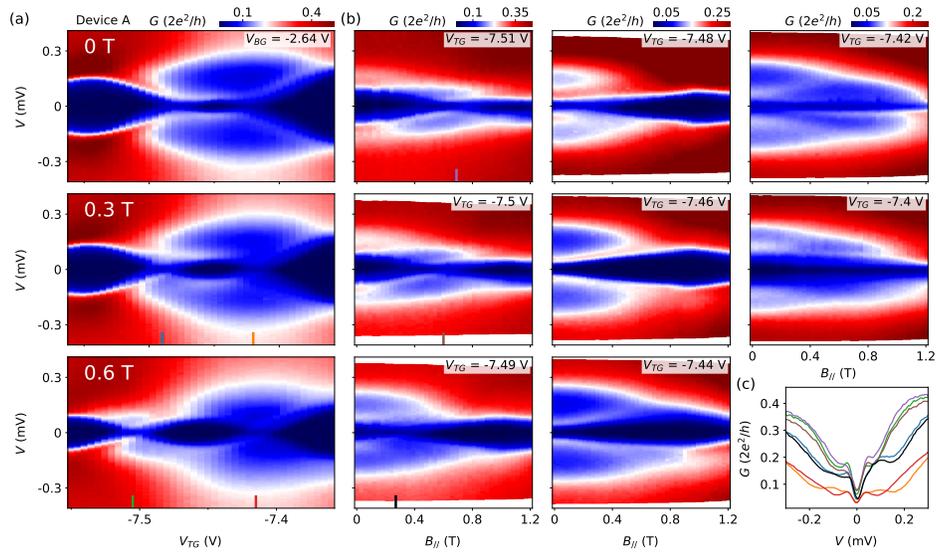

Supplement Fig. 6: Additional scans of the ABS in Fig. 2d-f. (a) $V_{TG}$ scans at $B = 0$, 0.3 and 0.6 T. (b) $B$-scans ($V_{TG}$ labeled). (c) Line-cuts of suppressed ZBPs from (a-b) labeled with corresponding color bars. $T ∼ 20$ mK

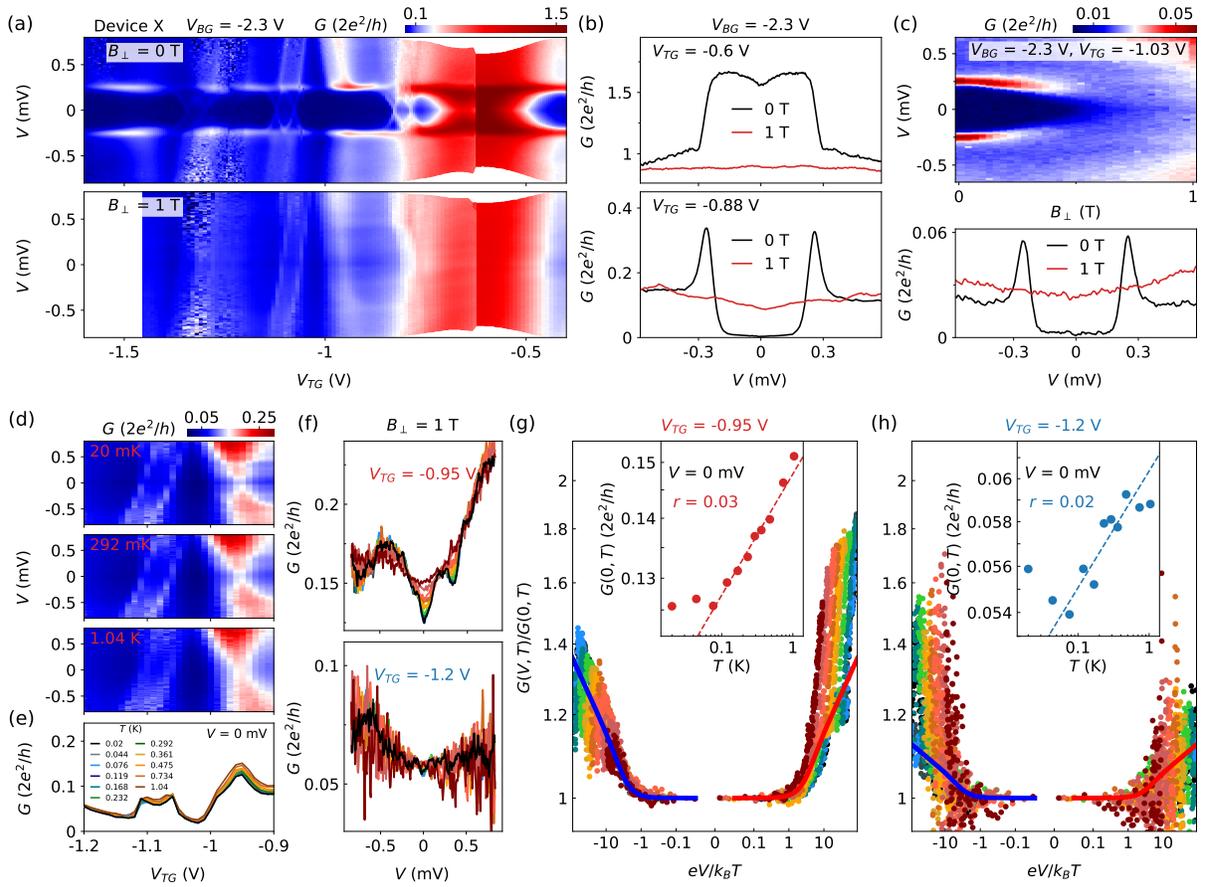

Supplement Fig. 7: Basic transport characterization of control Device X, a regular InAs-Al nanowire device without the dissipative resistor. (a) $G$ vs $V_{TG}$ and $V$ at $B = 0$ T (top) and 1 T (bottom). $B_\perp$ is in-plane and perpendicular to the nanowire. The top panel shows clear superconducting behavior with a hard gap in the tunneling regime (lower panel b) and Andreev enhancement in the open regime (upper panel b). The gap and enhancement features are suppressed at $B_\perp = 1$ T as shown in (b). The unintentional quantum dots and corresponding Coulomb blockades remain at 1 T. In the normal state (lower panel a), ECB suppression is either absent or barely visible: one may argue the red curve in the lower panel b as an extremely weak ECB suppression. Device X forms a sharp contrast with Device A-C where the ECB suppression exists in almost the entire gate space (Sfig. 2). (c) $B$-dependence of the gap: after gap closing the conductance curve is flat without obvious ECB suppression (lower panel). $T \sim 20$ mK for (a-c). (d) $T$-dependence of Device X in the normal state ($B_\perp = 1$ T), only three $T$ panels are shown for clarity. (e) Zero-bias line-cuts of the $T$-dependence do not show significant change over $T$, qualitatively different from Device A-C where the change is a factor of few or more than a decade. (f) $T$-dependence of two line-cuts from (d) and their 'power-law' fittings in (g) and (h) (inset for zero bias $G$). The 'power-law' does not show fitting for bias ($V$). For zero-bias $T$-dependence (insets), the change of $G$ is too small (only a fraction of the absolute value) to fit power-law. The change is in fact almost comparable to our measurement noise (inset of (h)). If a (unreliable) fitting was applied (dashed lines in insets), the extracted exponent $r \sim 0.02$ - $0.03$ translates to an effective dissipative resistance of a few hundred Ohms, possibly due to residual dissipaiton contributed by the Ti/Au contacts. Based on all the characterizations above, we conclude that our control Device X does not show clear ECB suppression with power-laws and can be treated as a no-dissipation or extremely weak dissipation device.

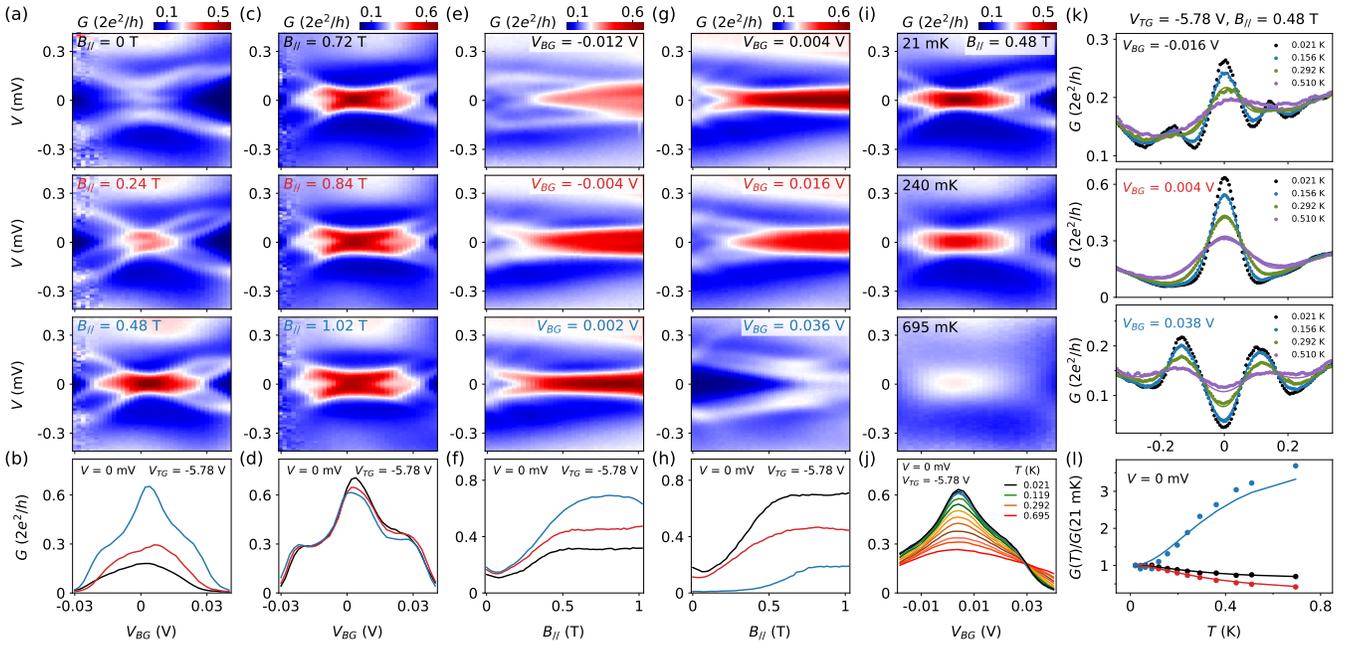

Supplement Fig. 8: Additional gate-scans, $B$-scans and $T$-dependence of the ZBP data set (Fig. 2g-h and Fig. 3e) in control Device X. (a-d) Gate-scans at fixed $B$s (labeled) with the zero-bias line-cuts shown in (b) and (d). (e-h) $B$-scans at different fixed gate voltages (labeled) with the zero-bias line-cuts shown in (f) and (h). The zero-bias line-cuts in $B$-scans resolve several 'plateau' features at non-quantized values. On these 'plateaus', further gate scans, see (b) and (d), do not resolve any plateau for these ZBPs. Therefore, the $B$-scan 'plateaus' are not real plateaus: a real plateau should be a plateau for all relevant parameter scans ($B$ and gate). (i) $T$-dependence of the ZBP by repeating the same 2D measurement at different $T$s (only three shown for clarity). The $T$-dependence in main text Fig. 3e corresponds to a line-cut from (i). (j) Zero-bias line-cuts for all $T$s. (k) $T$-dependence of additional line-cuts: the top two panels are for ZBPs while the lower one is for split-peaks corresponding to finite energy ABS. Only four $T$s are shown (as colored dots) for clarity. The solid lines (with corresponding colors) are thermal averaging simulation which shows reasonable match the ZBP data, and some deviations for finite energy ABS, possibly due to gap softening at high $T$. (l) $T$-dependence of the normalized zero-bias conductance for the three panels in (k) and its comparison with thermal averaging simulations (solid lines).

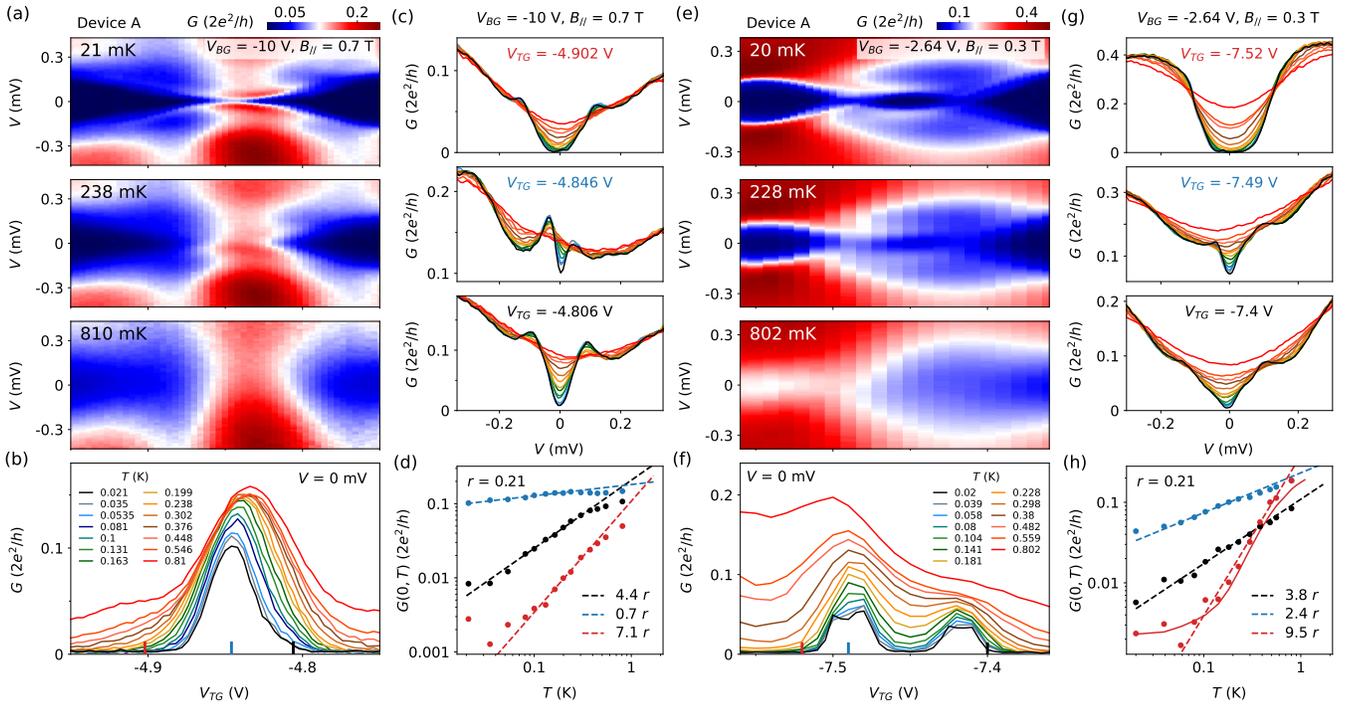

Supplement Fig. 9: Additional $T$-dependence of the ABS in Fig. 2. (a) corresponds to the ABS in Fig. 2a-c. $T$ varies from 21 mK to 810 mK (only three are plotted for clarity). (b) Zero-bias line-cuts of the $T$-dependence. (c) $T$-dependence of line-cuts at three different $V_{TG}$s, labeled by color bars in (b). (d) Power-law-like fits of the zero-bias $G$ for these three curves with the extracted exponents labeled, assuming $r = 0.21$. (e-h) Additional data for the $T$-depenence of the ABS in Fig. 2d-f and Fig. 3a-c. For illustrating purpose, we also plot the thermal averaging simulation ('convolution of Fermi distribution derivative') as the red line in (h), corresponding to the curves in the top panel (g): no obvious ABS or high energy ABS. The fitting between thermal simulation (red line) and power-law-like (red dashed line) are equally poor. This occasionally happens for curves like the top panel (g) without any zero or near zero-energy ABS. In this case, the zero-bias $G$ is probably dominated by Coulomb blockade and affected by gap softening at high $T$. Nevertheless, we restrict our 'thermal simulation' tool to only zero or near-zero energy ABS: in Device X (without dissipation), reasonably good match can be found between thermal simulation and $T$-dependence of ZBPs while in Device A (with strong dissipation), thermal simulation shows significant deviations from $T$-dependence of the near-zero energy ABS, as shown in Fig. 3b and 3d.

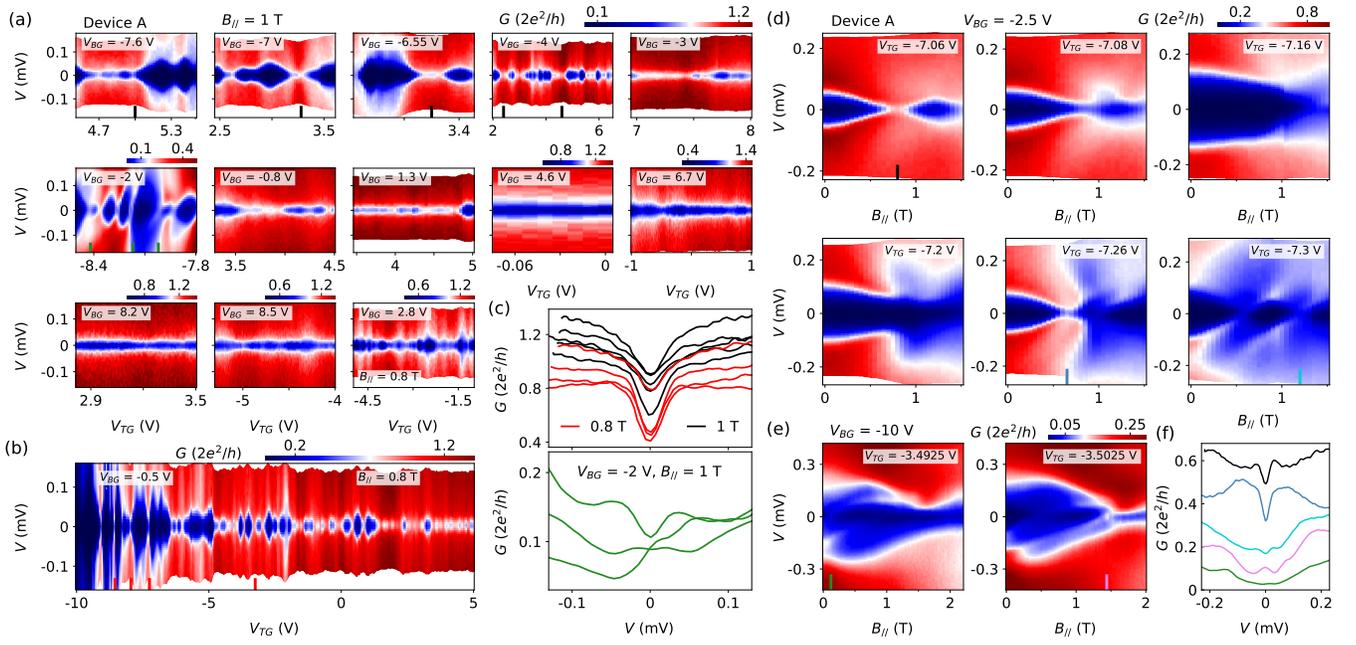

Supplement Fig. 10: (a-b) More gate scans in Device A at parallel $B$ of 1 T and 0.8 T (labeled). No ZBPs are found. (c) Line-cuts from (a-b), see labeled bars. (d-e) $B$-scans in Device A at different gate settings to resolve ZBP suppression. (f) Line-cuts from (d-e) where the pink curve with a faint ZBP-like feature is the only 'ZBP' we found in Device A after extensive searches. $T \sim 20$ mK.



\end{document}